\documentstyle[prb,multicol,aps,epsfig]{revtex}

\begin{document}

\title{Hopping versus bulk conductivity 
in transparent oxides: 12CaO$\cdot$7Al$_2$O$_3$}

\author{J.~E. Medvedeva and A.~J. Freeman}

\address{
Department of Physics and Astronomy, Northwestern,
Evanston, Illinois 60208-3112
}

\maketitle

\begin{abstract}
First-principles calculations of the mayenite-based oxide,
[Ca$_{12}$Al$_{14}$O$_{32}$]$^{2+}$(2e$^-$), reveal 
the mechanism responsible for its high conductivity.
A detailed comparison of the electronic and optical properties 
of this material with those of the recently discovered novel transparent
conducting oxide, H-doped UV-activated Ca$_{12}$Al$_{14}$O$_{33}$,
allowed us to conclude that the enhanced conductivity in 
[Ca$_{12}$Al$_{14}$O$_{32}$]$^{2+}$(2e$^-$) is achieved 
by elimination of the Coulomb blocade of the charge carriers.
This results in a transition from variable range hopping
behavior with a Coulomb gap in H-doped UV-irradiated Ca$_{12}$Al$_{14}$O$_{33}$
to bulk conductivity in [Ca$_{12}$Al$_{14}$O$_{32}$]$^{2+}$(2e$^-$).
Further, the high degree of the delocalization of the conduction 
electrons obtained in [Ca$_{12}$Al$_{14}$O$_{32}$]$^{2+}$(2e$^-$)
indicate that it cannot be classified as an electride, originally suggested.
\end{abstract}

\begin{multicols}{2}

Recently, an insulator-conductor conversion was discovered in 
a transparent oxide, 12CaO$\cdot$7Al$_2$O$_3$ or mayenite: 
after hydrogen annealing followed by ultraviolet (UV) irradiation, 
the conductivity of the material increases by 10 orders of 
magnitude \cite{Hayashi2002}. More recently, a further 100-fold enhancement of 
the conductivity (up to 100~S~cm$^{-1}$) was reported for 
[Ca$_{12}$Al$_{14}$O$_{32}$]$^{2+}$(2e$^-$) -- a material 
that was classified as a novel inorganic electride 
\cite{Matsuishi2003,Dye2003,Sushko}.
The technological importance of these two discoveries --
including a wide range of optoelectronic applications
\cite{Hayashi2002,Matsuishi2003}, the environmental advantages and 
the abundance of the ceramic constituents -- has stimulated 
enormous interest in materials of the mayenite family.
In this Letter, based on a detailed comparison of the electronic 
properties of the H-doped UV-irradiated Ca$_{12}$Al$_{14}$O$_{33}$ 
(abbreviated as C12A7:H$^0$) and 
[Ca$_{12}$Al$_{14}$O$_{32}$]$^{2+}$(2e$^-$) (abbreviated as C12A7:2e$^-$),
we demonstrate that despite chemical similarities of these two 
mayenite-based materials, their transport properties 
are not only quantitatively but also {\it qualitatively} different --
in C12A7:H$^0$ the conductivity is provided by variable range hopping,
while C12A7:2e$^-$ shows bulk conductivity.
In addition, the calculated charge density distribution for C12A7:2e$^-$
clearly shows high delocalization of the ``excess'' electrons -- a significant
property \cite{Dye97} that excludes this material from electride classification 
originally suggested \cite{Matsuishi2003}.

\begin{figure}
\centerline{
\includegraphics[width=6.4cm]{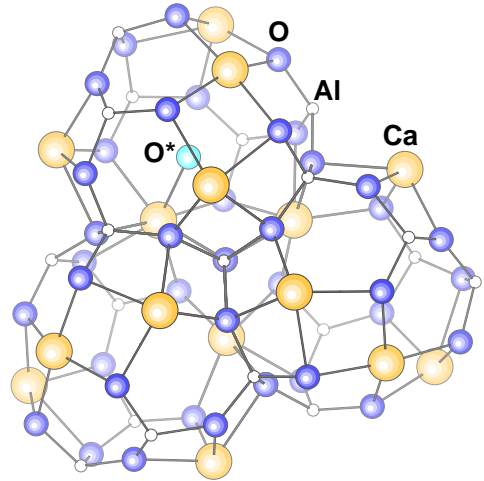}
}
\caption{Three of the 12 cages constituting the unit cell of mayenite, 
12CaO$\cdot$7Al$_2$O$_3$.
Located inside cages, the O$^{2-}$ anions (abbreviated as O$^*$) 
have irregular six-fold coordination with Ca atoms and 
are found (Ref. 8) to form a bcc lattice with
the crystallographic basis oriented randomly with respect
to that of the whole crystal.}
\label{str}
\end{figure}

The characteristic feature of mayenite, Ca$_{12}$Al$_{14}$O$_{33}$, 
is its nanoporous zeolite-like structure \cite{structure}: 
the unit cell with two formula units contains 12 cages 
of minimal diameter 5.6 \AA \, across (cf., Fig. \ref{str}). This framework
includes 64 of the oxygen atoms; the remaining two O$^{2-}$ ions 
(abbreviated as O$^*$ hereafter) provide charge neutrality 
\cite{Imlach71} and are located inside the cages, Fig. \ref{str}. 
The large cage `entrances' (about 3.7 \AA \, in diameter)  
make it possible to incorporate H$^-$ ions into the cages
according to the chemical reaction: 
O$^{2-}$(cage) + H$_2$(atm.) $\rightarrow$ OH$^-$(cage) 
+ H$^-$(another cage). The H-doped system remains an insulator 
-- until UV-irradiation is performed. 
First-principles calculations have shown \cite{origin} that 
in C12A7:H$^0$, the charge transport associated with the electrons 
excited off the H$^-$ ions occurs by electron hopping 
through the states of the H$^0$ and OH$^-$ (located inside the cages) 
and their Ca neighbors (belonging to the cage wall).
The detailed knowledge of the transport mechanism obtained for C12A7:H$^0$
allowed prediction of ways to enhance the conductivity by increasing 
the concentration of hopping centers (such as encaged H$^0$ or OH$^-$) 
which would provide additional hopping paths for the carrier migration
\cite{origin}.
In striking contrast, whereas in C12A7:2e$^-$ the encaged O$^*$ ions are removed
\cite{Matsuishi2003}, i.e., all cages are empty, the conductivity 
is increased by 2 orders of magnitude as compared to C12A7:H$^0$.

\vspace{0.2cm}

To understand the mechanism of the conductivity in C12A7:2e$^-$ 
and to determine the differences in the transport properties of these
two mayenite-based compounds, we performed first-principles electronic 
band structure calculations of mayenite with oxygen vacancies 
inside the cages (i.e., with all O$^*$ removed) and with 2e$^-$ added 
as a uniform background charge to provide charge neutrality.
Self-consistent solutions of the effective one-electron 
Kohn-Sham equations were obtained by means of the linear muffin-tin orbital 
method in the atomic sphere approximation \cite{asa}
for the cell of C12A7:2e$^-$ with one formula unit (a total of
58 atoms per unit cell which constitute six cages) and 
with periodic boundary conditions. In addition, 86 empty sheres were 
included to fill out the open spaces of the system \cite{Keller71}.

\begin{figure}
\includegraphics[width=8.5cm]{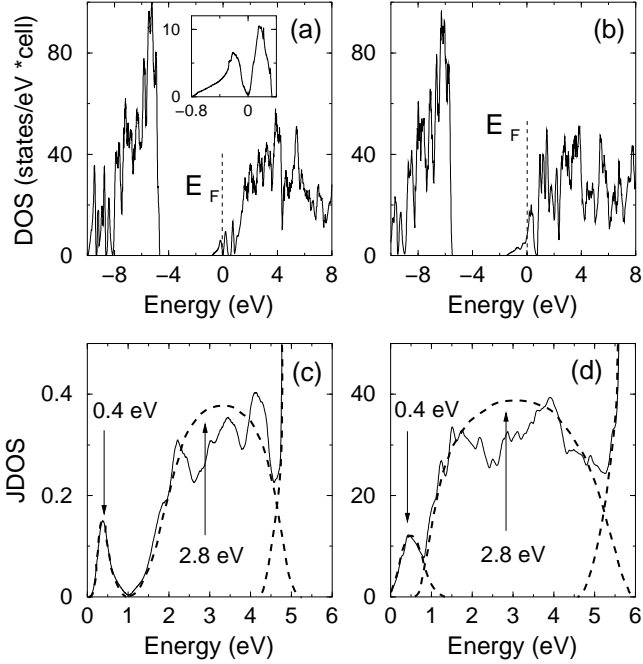}
\caption{
Total density of states of (a) C12A7:H$^0$ and the enlarged DOS near E$_F$
in the inset which shows the predicted Coulomb gap, 
and (b) C12A7:2e$^-$; the Fermi level is at 0 eV.
The joint DOS (solid lines) for (c) C12A7:H$^0$ and (d) C12A7:2e$^-$;
dashed lines are guide to the eyes; energy values and arrows show 
the positions of the observed absorption peaks 
(Ref. 1,2).
}
\label{dos}
\end{figure}

The calculated electronic densities of states (DOS) for 
Ca12Al7:H$^0$ and Ca12A7:2e$^-$ are found to be similar, 
cf., Fig. \ref{dos}(a) and \ref{dos}(b): in both cases, oxygen 2p states 
form the top of the valence band, Ca 3d states form the bottom of 
the conduction band and a hybrid band in the band gap crosses the Fermi 
level making both systems conducting. 
Using the optical selection rules, we calculated the joint DOS
that determines the positions of the characteristic absorption bands,
Fig. \ref{dos}(c) and \ref{dos}(d). As one can see, the overall
structure of the absorption spectra is similar for the two systems, 
and is in good agreement with experiment \cite{Hayashi2002,Matsuishi2003}. 
We find: (i) in C12A7:H$^0$, the 0.4 eV absorption peak is 
narrow, well-defined and separated from the 2.8 eV peak which 
corresponds to transitions from the occupied states of 
the hybrid band to the conduction band; (ii) by contrast, 
in C12A7:2e$^-$ the pronounced DOS at E$_F$ gives the non-zero 
absorption in the large wavelength limit; in addition, due to 
the increased width of the hybrid band, the absorption band 
centered at 0.4 eV substantially overlaps with the one at 2.8 eV, 
resulting in the observed black coloration of the samples \cite{Matsuishi2003}.

\begin{figure}
\includegraphics[width=4.2cm,angle=90]{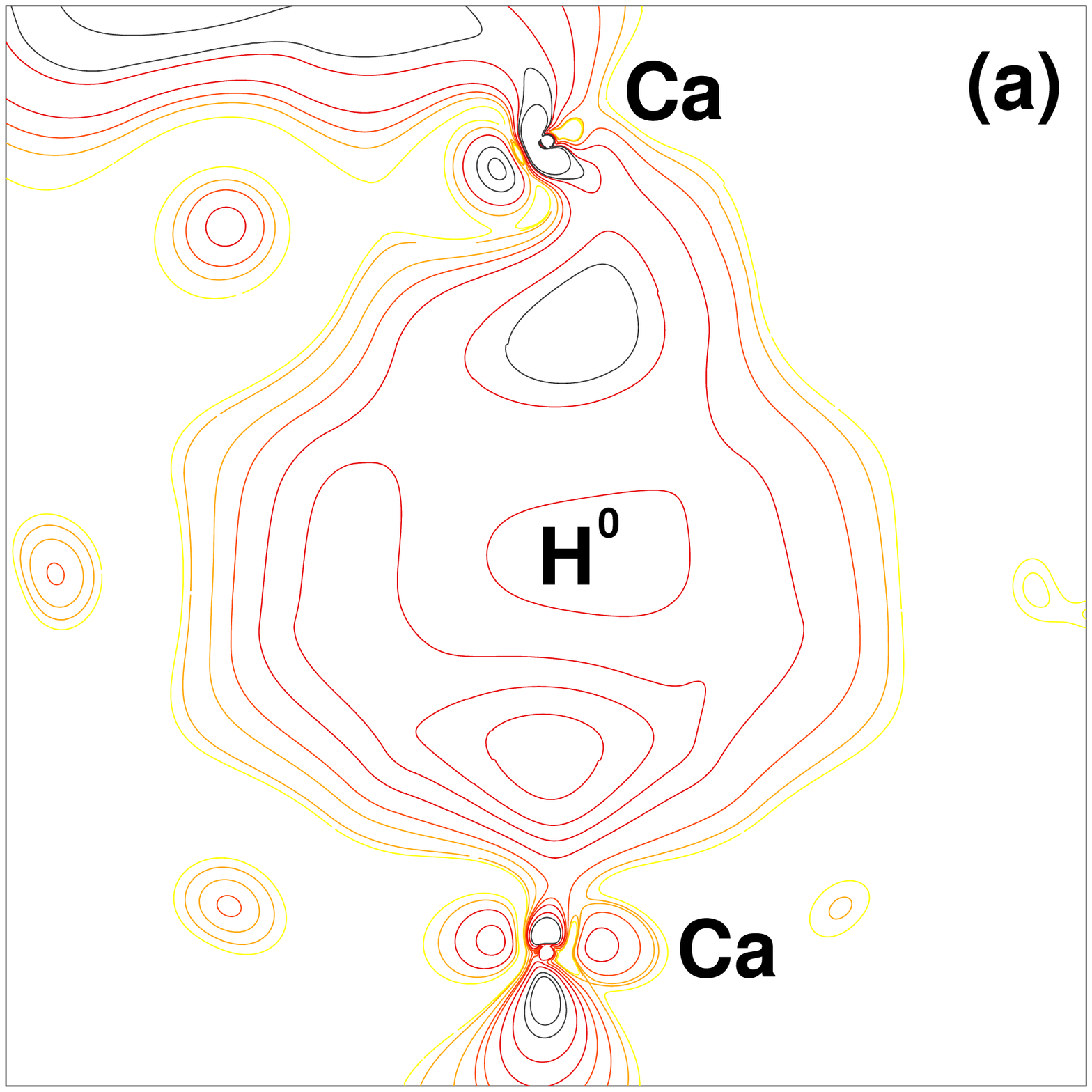}
\includegraphics[width=4.2cm,angle=90]{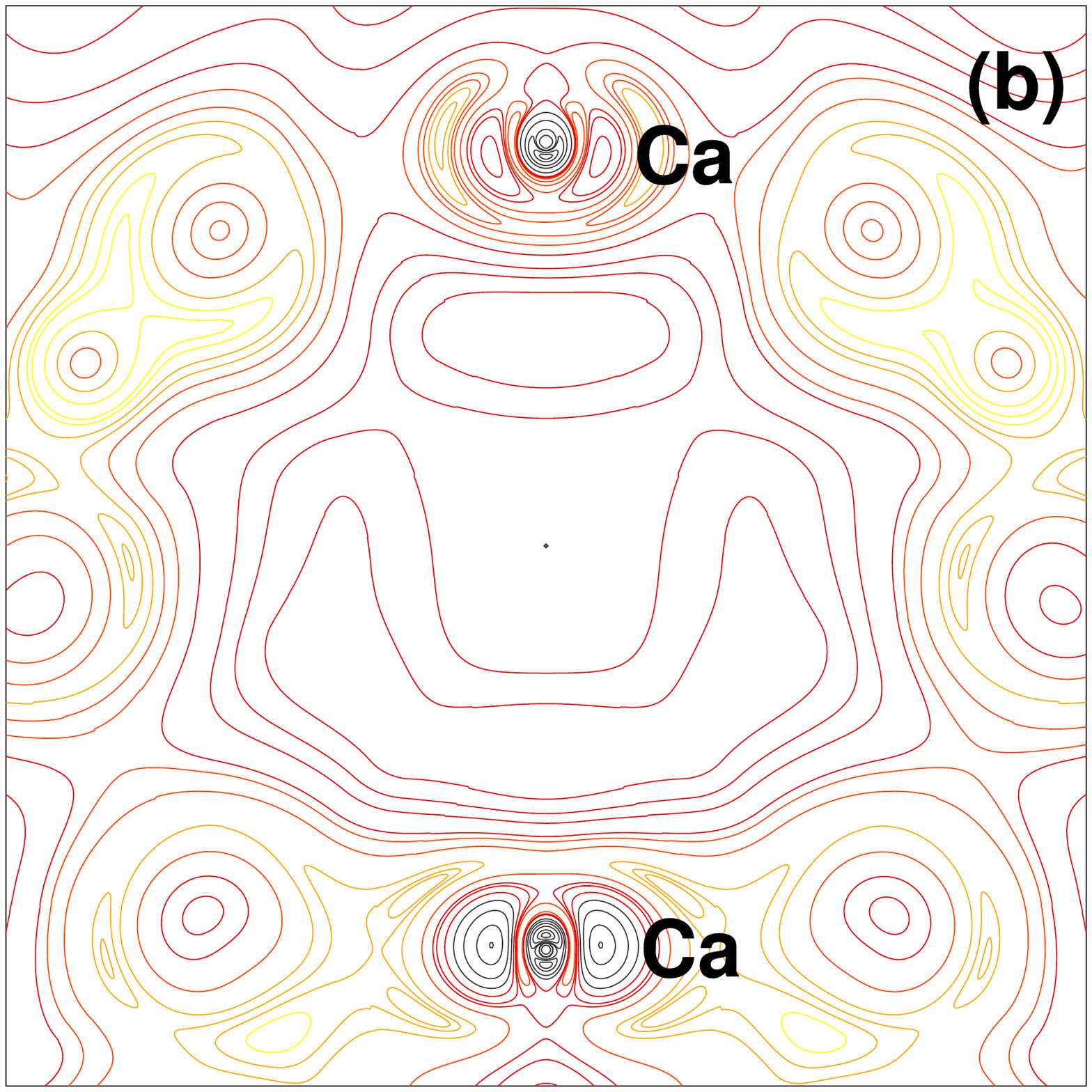}
\caption{
Contour map of the charge density distribution within a slice
passing through the center of (a) a cage with H$^0$ in C12A7:H$^0$
and (b) an empty cage (vacancy) in C12A7:2e$^-$.}
\label{charge}
\end{figure}

Despite the apparent similarity of the characteristic optical absorption 
peaks, we found that the conductivity mechanism in these two mayenite-based 
compounds is qualitatively different. This is clearly seen from 
a comparison of the
charge density distributions calculated in a 25 meV energy window
below E$_F$, cf., Fig. \ref{charge}. In the case of C12A7:H$^0$, 
the connected charge density maxima along the hopping path demonstrates
the hopping nature of the conductivity \cite{origin}.
In contrast to this trapping of the electrons on particular atoms, 
in C12A7:2e$^-$ the conduction electrons are found to be highly delocalized. 
Their extended wave function suggests a bulk mechanism for the conductivity.
Indeed, in C12A7:2e$^-$, all Ca and the encaged empty spheres
give comparable contributions to the DOS in the vicinity of E$_F$ 
(on average, 3.1 and 2.6 \%, respectively),
while in C12A7:H$^0$, only four of the 12 Ca atoms, i.e., the neighbors
of the encaged H$^0$ and OH$^-$, contribute significantly to the DOS 
near E$_F$ \cite{origin}.

The above description of the transport properties in C12A7:H$^0$ and 
C12A7:2e$^-$ differs from the original interpretation 
that was based on the formation of F$^+$ centers (i.e., 
electron trapping on the oxygen vacancy) inside positively 
charged cages \cite{Hayashi2002,Matsuishi2003,Sushko}. 
In the framework of the F$^+$ model, the released 
electron could hop between empty cages (there are eight of them per unit cell)
which strongly contradicts experimental observations for C12A7:H$^0$ 
that the area not exposed to UV irradiation remains insulating 
\cite{Hayashi2002,Bertoni}. In sharp contrast to the F$^+$ model, 
our calculations \cite{origin} demonstrate clearly that the conducting channels
do not involve empty cages and are formed only in the UV-irradiated area 
because the untreated H$^-$ ions prevent the creation of such hopping paths
in the light protected region -- in agreement with experiment
\cite{Hayashi2002,Bertoni}.

Consistent with this view, we found that in C12A7:2e$^-$
the conduction electrons are not localized inside the cages, cf. Fig. \ref{charge}(b).
This indicates that C12A7:2e$^-$ cannot be considered to be an electride 
\cite{Matsuishi2003,Dye2003,Sushko} which is essentially classifies 
those compounds in which 
the electron density is confined within structural cavities and 
tends to avoid the regions occupied by cations \cite{Dye97,Li2003}.
Our result is in contrast with the first theoretical study of C12A7:2e$^-$
that was based on the embedded defect cluster approach \cite{Sushko} which 
excluded possibilities other than electron localization inside the cages,
i.e., formation of the F$^+$ center;
strikingly, the resulting schematic picture of the energy levels
(cf., Fig. 3 of the Ref. \onlinecite{Sushko}) disagreed with 
the observed optical absorption energies \cite{Hayashi2002}.

Thus, our first-principles band structure calculations provide 
a clear physical explanation of the 100-fold enhancement of the conductivity 
in C12A7:2e$^-$ as compared to C12A7:H$^0$.
In the H-doped case, the released electrons migrate along a well-defined channel 
-- the hopping path \cite{origin}. Moving one-by-one, they interact (repel) 
strongly with each other resulting in the formation of the parabolic 
Coulomb gap in the DOS at E$_F$ (cf., Fig. \ref{dos}(a) inset).
The value of the gap can be estimated as $e^2/\epsilon \, r_{ch}$, where
$e$ is the elementary charge, $\epsilon$ is the dielectric constant, and 
the characteristic distance between two electrons in the unit cell, 
$r_{ch}$, is equal to the cube root of the unit cell volume divided by 2. 
Using the lattice parameter of cubic mayenite, 11.98 \AA, 
and $\epsilon$=2.56 (Ref. \onlinecite{Zhmoidin}), yields 0.3 eV for the value
of the Coulomb gap
which agrees well with the calculated \cite{origin} splitting of 
the hybrid band, 0.38 eV, and with the observed \cite{Hayashi2002}
absorption peak at 0.4 eV.

This successful interpretation of the transport properties in C12A7:H$^0$
suggests a way to improve the conductivity: an increased number of 
transport channels should alleviate the electron repulsion.
Consistent with this view, a 100-fold increase of the conductivity 
is observed for C12A7:2e$^-$, 
where, loosely speaking, the number of conducting channels can be 
considered very large so that the injected electrons have more freedom 
and the ``Coulomb blockade'' does not occur.
Following this interpretation, we predict that even higher 
conductivity can be obtained in similarly treated 
Si-substituted mayenite, [Ca$_{12}$Al$_{10}$Si$_4$O$_{32}$]$^{6+}$(6e$^-$), 
where two additional oxygen vacancies inside the cages 
would result in a three-fold increase of the carrier concentration
(up to 7$\times$10$^{21}$~cm$^{-3}$) in this material \cite{tobe}.

Finally, we stress that the enhanced conductivity in C12A7:2e$^-$ comes 
at the cost of greatly increased optical absorption. The latter occurs due to
the increase of (i) the density of states of the hybrid band in the band gap
and (ii) the width of this hybrid band -- resulting in intense transitions
from the occupied states of the hybrid band to the conduction band
in the visible range which excludes 
the possibility of using this oxide as a transparent conducting material.
Fortunately, the details of the electronic band structure and 
the conductivity mechanism found here will help in the search 
for new candidates with improved optical and transport properties.

We thank M.~I. Bertoni and T.~O. Mason for close collaboration
in carrying out suggested conductivity measurements for the UV-light protected 
C12A7:H$^0$ samples.
Work supported by the DOE (grant N DE-FG02-88ER45372)
and the NSF through its MRSEC program at the Northwestern University 
Materials Research Center. Computational resources have been provided by
the DOE supported NERSC.


\end{multicols}
\end{document}